\newif\iflayout
\layouttrue

\iflayout
   \documentclass[twocolumn,superscriptaddress,prl]{revtex4}
\else
   \documentclass[preprint,superscriptaddress,prl]{revtex4}
\fi

\usepackage{epsfig,amssymb}
\usepackage[latin1]{inputenc}

\begin{document}

\title{Coexistence of ferro- and antiferromagnetic order in Mn-doped Ni$_2$MnGa}
\author{J. Enkovaara}
\email{jen@fyslab.hut.fi}
\affiliation{Laboratory of Physics, P.O. Box 1100, Helsinki University of
Technology, FIN-02015 HUT, Finland}
\author{O. Heczko}
\affiliation{Laboratory of Biomedical Engineering, P.O. Box 2200, Helsinki
University of Technology, FIN-02015 HUT, Finland}
\author{A. Ayuela}
\author{R. M. Nieminen} 
\affiliation{Laboratory of Physics, P.O. Box 1100, Helsinki University of Technology, FIN-02015 HUT, Finland}

\begin{abstract}
Ni-Mn-Ga is interesting as a prototype 
of a magnetic shape-memory alloy showing large magnetic field induced strains.  
We present here results for the magnetic ordering of Mn-rich Ni-Mn-Ga alloys 
based on both experiments and theory.
Experimental trends for the composition dependence of the magnetization are
measured by a vibrating sample magnetometer (VSM)
in magnetic fields of up to several tesla and at low temperatures.
The saturation magnetization 
has a maximum
near the stoichiometric composition and it decreases with increasing Mn content.
This unexpected behaviour is interpreted via first-principles calculations 
within the density-functional theory.
We show that extra Mn atoms are antiferromagnetically aligned to the other
moments, which explains the dependence of the magnetization on composition.
In addition, the effect of Mn doping on the stabilization of the structural
phases and on the magnetic anisotropy energy is demonstrated.
\end{abstract}


\maketitle

\begin{figure}[t] 
\centering
\epsfig{file=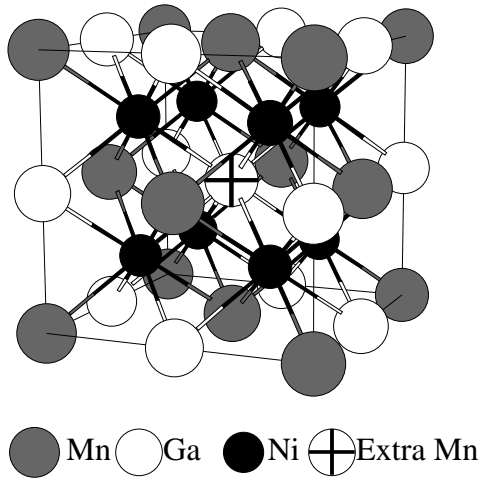,width=.7\columnwidth,keepaspectratio=true}
\caption{L2$_1$ supercell of Ni$_2$Mn$_{1.25}$Ga$_{0.75}$.}
\label{struct}
\end{figure}

Ni-Mn-Ga alloys have unique magnetoelastic and magnetoplastic properties,
 which  make them highly interesting for many novel technological applications 
 such as smart actuator materials \cite{nature,aplications}.
The material properties are sensitive to the actual composition \cite{old}. 
The best alloys to date are rich in Mn  and poor in Ga with respect to the 
perfect Ni$_2$MnGa  stoichiometry.
In particular, a shape memory alloy around Ni$_{2}$Mn$_{1.25}$Ga$_{0.75}$
composition shows excellent  
magnetoelastic properties with  strains of up to 10 \% under moderate
magnetic fields. 
Its recent discovery has aroused substantial experimental and theoretical
activity \cite{interest}.
Typically these materials are produced by several techniques, such as
directional 
solidification and Bridgman growth \cite{working}, and  one
has also been
able  to grow them epitaxially on GaAs substrates \cite{substrate}. 
Apart
from the composition dependence, the ordering of the atoms in the different
sublattices 
within the L2$_1$  structure, as depicted in Fig.~\ref{struct}, is not yet well
understood. 
A thorough  theoretical investigation of the phase equilibrium
is also lacking, and many details about bonding and ordering in
this system are not yet well understood \cite{ordering}.

In order to facilitate a technological breakthrough, the magnetic
properties of this material should be explored in detail.
For the perfect stoichiometric case, experimental as well as
theoretical works agree on the saturation  magnetization of 4.1 $\mu_B$ per
formula unit \cite{stoichiometry,alloys,variants}. 
For Mn-rich alloys, it
is expected that extra Mn atoms substitute Ga atoms in the Ga sublattice.
The Mn atoms would contribute by 3.5 $\mu_B$ to the total magnetic moment
as described in Refs. \cite{stoichiometry,alloys,variants}. 
However, addition of Mn does not correlate with an increase in the overall
magnetization, as shown in the experimental results  
below.  Thus there is a serious difference between the experimental results
and the expected trends. 
In this Letter we address this problem
by a combined experimental and theoretical study.  
 
Most of the alloys used to date have been developed with a metallurgical
approach. 
In contrast,
we use alloy theory in order to identify the phases \cite{variants}, 
to understand the martensitic phase transition mechanism
\cite{variants,anisotropy+other}, and 
to predict the possible candidates for future materials \cite{alloys}.
Here we show a nearly perfect agreement between theory and experiment 
considering magnetic order in alloying with Mn atoms.
As a bonus we also provide a microscopic explanation for further stabilization
of the  martensite phases appearing at low temperature. Finally, we report the
calculated 
magnetic anisotropy energy
which deviates from the results based on the average electronic
concentration.  

\begin{figure}[t] 
\centering
\epsfig{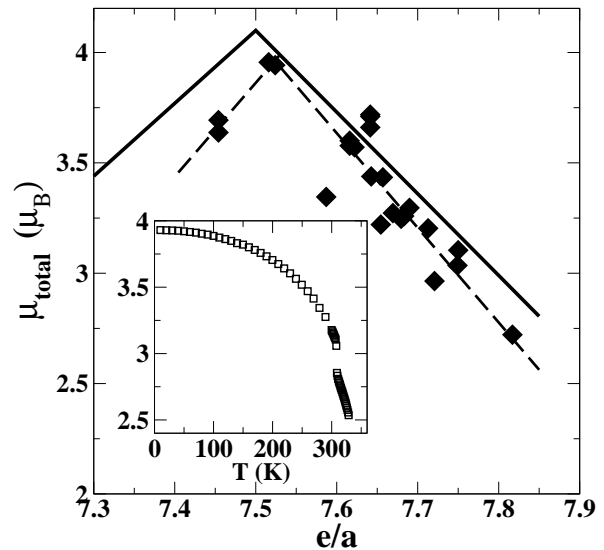}
\caption{Saturation magnetization vs the average number of 
valence electrons per atom (e/a). The dashed line is a linear fit to the
experimental data and the solid line is the theoretical prediction.
The inset shows the temperature dependence of the magnetization.}
\label{exp}
\end{figure}

The details of the experiments are the following.
Various alloys with compositions close to stoichiometric Ni$_2$MnGa were melted
from pure elements and nearly single-crystal ingots were produced by a
modified Bridgman method. The as-cast ingots were
homogenized at 1000~$^\circ$C. 
The saturation magnetization is studied as a function of the composition.
The chemical composition of the samples is determined
by energy dispersive spectroscopy (EDS). 
While the content of Ni remains as almost constant 50~\%, the content
of Mn spans the range between 19~\% and 34~\%.
The magnetization curves are
measured with a vibrating sample magnetometer (VSM) up to a field of 2T and with
a SQUID magnetometer up to a 5T field. Measurements are made for several
temperatures down to
10 K (see the inset in Fig~\ref{exp}).
The spontaneous saturation magnetization is determined by extrapolation to the
zero field limit from the magnetization measurements at 10~K.  

For modeling, we use  the generalized-gradient-approximation
\cite{another,lsd} of spin-density-functional 
theory. 
For the total energy calculations, we use both a full-potential method (FLAPW)
\cite{Wien97} and a pseudopotential scheme
\cite{vasp,newPAW}.
As shown in previous calculations, the Mn magnetic moment is high, 
and we have confirmed the pseudopotential validity by
comparing full potential calculations with pseudopotentials calculations.
In this particular case, the pseudopotential treatment  is valid as concerns
the magnetic ordering and accurate enough for dealing with the 
main features of the structural phase transitions in these materials.
The details of these results will be discussed elsewhere \cite{jussi}. 
The substitutional and magnetic alloying is treated by replacing one Ga
atom 
with a Mn atom in the 16-atom L2$_1$ supercell (Fig.~\ref{struct}), resulting in
the composition 
Ni$_2$Mn$_{1.25}$Ga$_{0.75}$.   
The magnetic moment of the extra Mn atom is allowed to have both parallel and
anti-parallel alignment with respect to the other Mn atoms, 
collinearly ordered, and the total energy is minimized with respect to each
magnetic configuration. 
   
Fig.~\ref{exp} shows the magnetization measurements of
Ni$_2$Mn$_{1+x}$Ga$_{1-x}$ alloys as a function of the average electron
concentration measured by the electron to atom ratio (e/a).  
The temperature dependence of the magnetization (see the inset in
Fig~\ref{exp}) has been 
measured at the field H= 2 T, which is larger than the saturation field.
The saturation magnetization shows a peak at 7.5 valence 
electrons/atom corresponding to the perfect stoichiometric Ni$_2$MnGa alloy.
The magnetic moment has a value of about 4 $\mu_B$ per formula unit which 
 is in agreement with previous experimental data \cite{stoichiometry} 
on stoichiometric Ni$_2$MnGa as well as with the first-principles calculations 
\cite{alloys,variants}.
The experimental data show that the
magnetic moment decreases when the electronic concentration increases over 7.5.
This finding is in agreement with the data in Ref. \cite{jin} which
have been collected from several sources and show smaller values for
the  magnetization since the saturation 
is not well guaranteed. This is because measurements are made with either a low
field or  
at high temperatures.

\begin{figure}[t] 
\centering
\epsfig{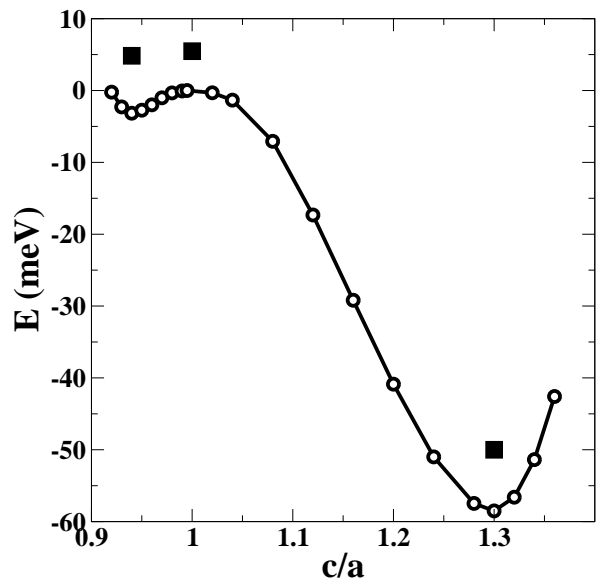}
\caption{Calculated total energy of Ni$_2$Mn$_{1.25}$Ga$_{0.75}$ per formula
unit as a function
of the tetragonal distortion $c/a$ ratio. Circles mark the antiferromagnetic
aligment of the extra Mn and squares mark the ferromagnetic alignment.}
\label{ene}
\end{figure}

In order to explain the experimental results above and to gain insight into
the precise magnetic structure, 
we present the following theoretical results. 
As a first step, let us analyse the magnetic properties based on the average 
electron concentration within the rigid
band approximation. It is found that the Mn moment remains constant but the Ni
moment decreases linearly with
increasing electron concentration. This picture 
cannot therefore 
explain the experimental decrease of the magnetization when lowering electron
concentration below 7.5. Also, the decrease of the magnetic moment per Ni atom
is only 
0.1~$\mu_B (e/a)^{-1}$  
which is smaller than in the experiments. 
When the extra Mn atom is included explicitly in the calculation, the
configuration with \emph{anti-parallel} alignment of Mn moments is 
energetically favourable as seen in Fig.~\ref{ene}.
An analysis of the local 
Mn magnetic moment gives a nearly constant value of around 3.5
$\mu_B$ regardless of the alignment, so that the total magnetic moment has a
value of 5.0 $\mu_B$ for 
the ferromagnetic configuration, and a value of 3.6 $\mu_B$ for the
antiferromagnetic one. The latter value is in good agreement with the
experiments.   

The variation of the magnetization with the composition can be estimated with
a simple model. As the extra Mn atoms couple antiferromagnetically with
the neighbouring Mn atoms, every additional Mn atom decreases the total magnetic
moment by 3.5 $\mu_B$. The total magnetic moment of Ni$_2$Mn$_{1+x}$Ga$_{1-x}$
is then given by 
$\mu_{total}=2 \mu_{Ni}+ (1-|x|) 3.5 \mu_{B}$
where the Ni moment $\mu_{Ni}$ is varied around the stoichiometric value
0.3~$\mu_B$ with the electron concentration
according to the rigid band results.
One should note that the variation of the Ni moment affects only slightly the
total moment whose variation is determined mostly by Mn.
The linear fit of the experimental data agrees well with the
line 
predicted by the theory, assuming a constant 50~\% content of Ni
\cite{error}. The 
reason for the peak of the saturation magnetization with electron
concentration around 7.5   
as well as the decreasing trend  
of the saturation magnetization is identified here in a natural way. We can
conclude that the Mn atoms substituted at the Ga sites are antiferromagnetically
coupled to the Mn atoms at Mn sites. This finding can be reasoned by the tendency
of close Mn atoms to favor an antiferromagnetic alignment,  
typical for several Mn compounds \cite{kubler}. 
The interesting new feature here is the coexistence of ferromagnetic and
antiferromagnetic Mn-Mn interactions in the same alloy.

It should be pointed out 
that during cooling the alloys undergo martensitic transformations, in some 
cases actually several intermartensitic transformations from so called 5M to
7M and to  
non-modulated martensite \cite{Oleg}.
For instance, in the inset of Fig.~\ref{exp}, an abrupt jump in the curve
indicates the austenite-martensite transformation at about 305 K.
During the 
transformation the saturation magnetization can change 10\%, 
but 
the low temperature moment remains nearly equal as shown in previous
calculations \cite{variants}. The antiferromagnetic alignment is
energetically favourable also in the martensitic phase (Fig.~\ref{ene}), so
that the phase 
transitions do not affect the above conclusions. 

We next discuss the existence of martensitic phases for the desirable 
compositions. In the stoichiometric alloy total energy calculations show two
energy minima with tetragonal structures $c/a\sim 0.94$ and $c/a\sim 1.3$
\cite{variants}. However, when 
the Mn-doping is modeled with the rigid band approximation the energy minimum
at $c/a\sim 0.94$ disappears. 
In Fig.~\ref{ene} we present total
energy calculations beyond the rigid band approximation where the additional
Mn is explicitly included. 
There is now a stable structure at $c/a\sim0.94$ both with parallel and
anti-parallel 
alignments of the extra Mn, but the antiferromagnetic coupling clearly stabilizes
 the minimum over the ferromagnetic case.  
When taking into account also orthorhombic distortions with the
antiferromagnetically aligned extra Mn, we find a new energy minimum with
$c/a\sim0.93$ and $b/a\sim0.97$.
The energy of this orthorhombic minimum is between the two tetragonal minima,
so that the theoretical order of the phases agrees with the experimental
findings \cite{Oleg}. 
We can conclude that the structural and magnetic ordering beyond
simple electronic averaging is important in stabilizing some of the phases.

Another key property in magnetism is the magnetic anisotropy energy. 
Previously the composition dependence of the magnetic anisotropy energy has
been analyzed by taking into account only the average electron concentration
\cite{anisotropy+other}. It is shown in Ref.~\onlinecite{anisotropy+other} that
the magnetic anisotropy energy 
decreases with increasing electron concentration in agreement with
experiments. Since the doping changes also the local structural and magnetic
properties we refine here the analysis and calculate the magnetic anisotropy
with the 
extra Mn explicitly included. With average electron concentration 
corresponding to Ni$_2$Mn$_{1.25}$Ga$_{0.75}$ composition the anisotropy
energy of the 
$c/a=0.94$ structure is 50~$\mu eV$ per formula unit. When including the extra
Mn explicitly the magnetic anisotropy energy is 150~$\mu eV$ for the
ferromagnetic alignment and 100~$\mu eV$ for the antiferromagnetic one, where
the last value is in good agreement with the experimental one of 90~$\mu eV$
\cite{Oleg2}. 
It is seen that even though simple averaging of electrons can reproduce the
correct trends, the structural and magnetic ordering is important for 
quantitative determination of the magnetic anisotropy energy.

In conclusion, the magnetic properties of Mn-doped Ni$_2$MnGa
alloys are measured for several 
concentrations and calculated for those showing interesting magnetoelastic
properties.  
Experimentally, we see that doping has a strong effect on the 
saturation magnetization.
The computational results within the density-functional theory
show that the extra Mn is antiferromagnetically aligned to the other atoms 
in the L2$_1$ lattice and 
a comparison of the experimental and the theoretical magnetizations
confirms this picture.
The ordering of the extra Mn is shown to affect also the appearance of the
tetragonal and orthorhombic structures and the magnetic anisotropy energy.
The main result of this  
work, the magnetic alignment of the extra Mn, could serve as an impetus
for further works on the ordering issue.
For example, the interaction of frustrated magnetic sublattices in Ni-Mn-Ga
should be revealed by using sublattice-sensitive probes.

This work has been supported by the Academy of Finland
(Centers of Excellence Program 2000-2005), by the National Technology
Agency of Finland (TEKES) and the consortium of Finnish companies (AdaptaMat
Oy, Metso Oyj, Outokumpu Research 
Oy). O. Heczko thanks Miroslav Marysko and Milos Jirsa from
Institute of Physics, Academy of Science, the Czech Republic for help with
the magnetization measurements. 
Computer facilities of the Center for
Scientific Computing (CSC) Finland are greatly acknowledged.
We thank A. Foster for the careful and critical reading of the manuscript.


\begin{thebibliography}{99}
\bibitem{error}
In reality the Ni content varied from 48 to 53 atom\%. Most of the scatter is,
however, due to errors in the chemical analysis. Also, the Mn moment may vary
slightly giving small variation from the exact linear behaviour.
\end{thebibliography}
\end{document}